\documentclass[journal]{IEEEtran}
\usepackage[utf8]{inputenc}
\usepackage{graphicx}
\usepackage[many]{tcolorbox}
\usepackage{natbib}
\bibpunct{(}{)}{,}{a}{}{;}
\begin{document}

\newcommand{\keywords}[1]{\par\addvspace\baselineskip
\noindent\keywordname\enspace\ignorespaces#1}

\title{Socially Intelligent Interfaces for Increased Energy Awareness in the Home}
\author{Jussi Karlgren, Lennart E. Fahl\'en, Anders Wallberg, Pär Hansson, \\ Olov Ståhl, Jonas Söderberg, Karl-Petter Åkesson \\
Swedish Institute of Computer Science (SICS) 
}

\maketitle

\begin{tcolorbox}[colback=red!10!white,
                     colframe=red!20!black,
                     title=\textsc{This paper has been published in the Proceedings of the First Conference on the Internet of Things},  
                     center, 
                     valign=top, 
                     halign=left,
                     before skip=0.8cm, 
                     after skip=1.2cm,
                     center title, 
                     width=3in]
In: Floerkemeier C., Langheinrich M., Fleisch E., Mattern F., Sarma S.E. (eds) The Internet of Things. Lecture Notes in Computer Science, vol 4952. Springer, Berlin, Heidelberg. https://doi.org/10.1007/978-3-540-78731-0\_17

  \end{tcolorbox}

\begin{abstract}
  This paper describes how home appliances might be enhanced to
  improve user awareness of energy usage.  Households wish to lead
  comfortable and manageable lives. Balancing this reasonable desire
  with the environmental and political goal of reducing electricity
  usage is a challenge that we claim is best met through the design of
  interfaces that allows users better control of their usage and
  unobtrusively informs them of the actions of their peers. A set of
  design principles along these lines is formulated in this paper. 
  We have built a fully functional prototype home appliance with a socially aware interface to signal the
  aggregate usage of the user's peer group according to these principles, and present the prototype in the paper.

\end{abstract}

\begin{figure*}[htbp]
\begin{center}
\includegraphics[ height=7cm]{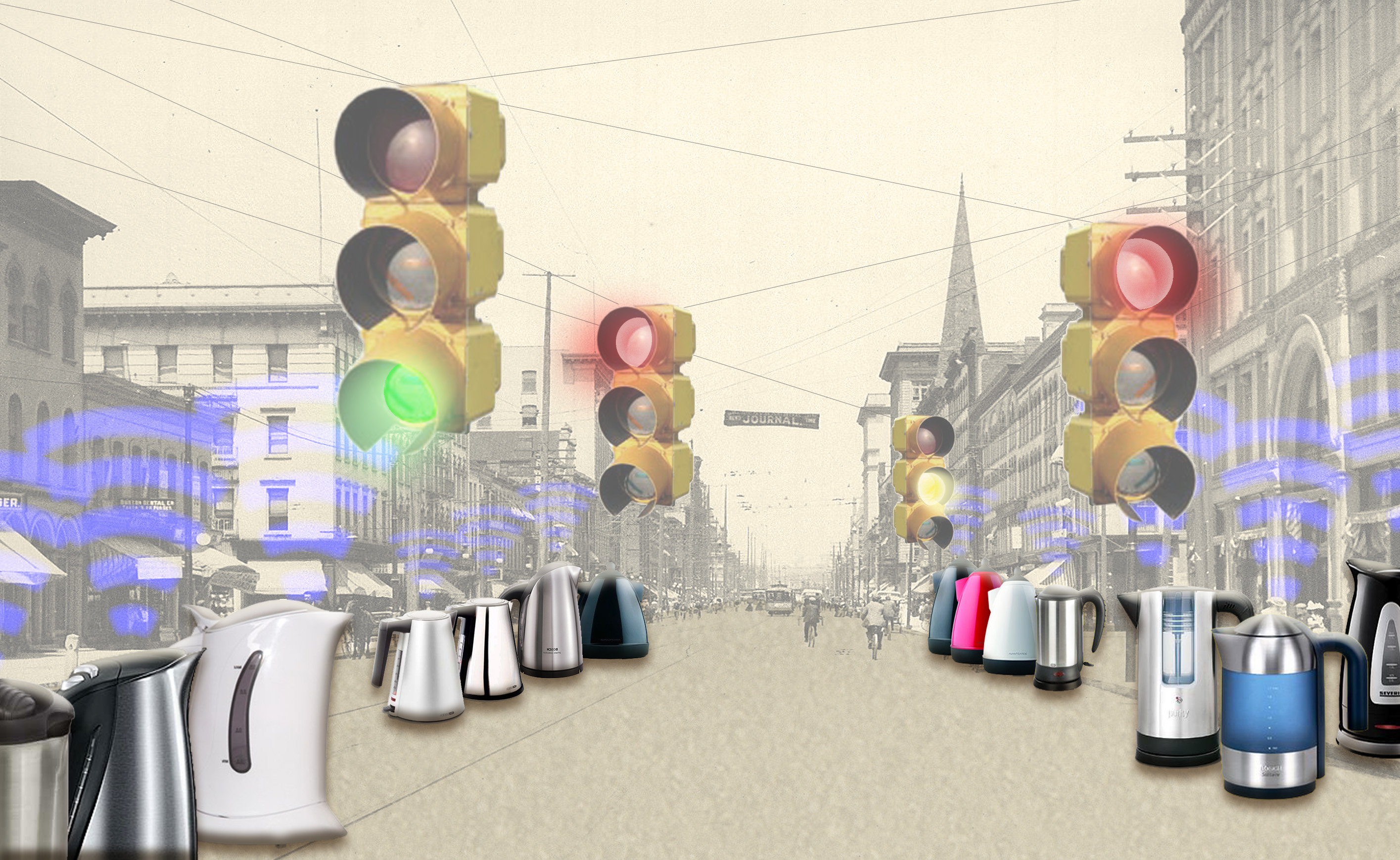} 
\end{center}
\caption{Communicating energy usage between peers.}\label{komm}
\end{figure*}

\section{Energy Usage in the Home --- an Interface Issue}

Monitoring and controlling energy usage is a small component in an
increasingly intellectually demanding everyday life. Many other
choices make demands on the schedule, budget, and attention of the
consumer. 

However, energy usage is becoming an increasingly important facet of
personal choice: the choices made by individual consumers have
consequences, both social and environmental. In public discourse,
energy usage is discussed as an arena where individual choice makes a
real and noticeable difference for the society and the ecological
system. This is a driving force and a motivating factor for the
individual consumer.

At the same time, another strand of development makes itself known in
the home and the household: the incremental digitalisation of home
services. The introduction of more and more capable entertainment,
media, and gaming appliances on the one hand, and more powerful
communication devices on the other, has brought typical homes to a
state where there is a high degree of computing power but very little
interconnection between appliances. It is easy to predict the future
but more difficult to get it right, and the history of ``intelligent
homes'' and other related research targets is littered with mistaken
projections and predictions~\cite{harper,smart}. However, it can safely be envisioned that
the future will hold better interconnection between the various
computing systems in the home, and that more digital services and
appliances for the home will be introduced apace. We believe that
systems for energy monitoring and control are an important vehicle for
home digitalisation and that they are a natural locus for the
convergence of the various functions foreseeable and present in a
home.

Our research method is to instantiate various combinations of design principles into fully functional product prototypes, some of which are used to prove a point in a demonstration, others which are field tested, and yet others which are used as a basis for industrial projects. The prototypes are meant to illustrate and test the principles - not to conclusively prove or disprove them. 

This paper begins by giving the points of departure for this concrete project, in terms of political and societal goals and in terms of consumer and user needs;it continues by outlining some central design principles taken into consideration during the design cycle; it then describes a concrete prototype for raising energy awareness and for providing unobtrusive mechanisms for better control of electrical energy usage in the home, instantiating the design principles in question.

\section{Goals}

Current societal and political attention in to a large extent focussed on questions of environmental sustainability. Chief among those question is that of energy usage and turnover. The operation, management, and maintenance of energy usage is to a large extent technological --- but based on the interaction between choices made to uphold a desired lifestyle, to conserve energy, participate responsibly in attaining societal objectives, and to preserve the integrity and habitability of one's personal everyday life. To this end, our work recognises a number of background non-technological goals. 

\subsection{First Goal: Reducing Electricity Usage}

An overriding political goal in Europe is to reduce energy usage. 
While energy is a raw material in several central industrial and infrastructural processes, as
well as a key resource in transport and in heating and cooling indoor locations, 
much of energy usage is incidental to the primary task at hand -- that of leading a comfortable life or performing services and producing goods in a comfortable environment. We will in this example focus on one aspect of energy usage, that of electricity usage in household environments.  

Household electricity usage has risen noticeably in the most
recent ten-year period~\cite{stem05}. There are opportunities to reduce electricity usage, inasmuch much of
the increase can be attributed to unaware electricity spill, caused by
standby systems or inadvertent power-up of home appliances.

The current public policy on energy matters includes
the goal to reverse this trend and to considerably reduce household electricity usage within the next decade. There are studies that show that in
spite of a higher density of kitchen appliances and a rapid and
increasing replacement rate as regards such appliances, kitchens use
less electricity today than ten years ago\cite{fyra}, which speaks towards the positive effect of
technological advancement and towards new technology being a solution,
not a hindrance to energy efficiency. Reduction in electric energy usage in households cannot be accomplished through
tariff manipulation, since the politically appropriate price range of
a unit of electricity cannot provide adequate market differentiation.

\subsection{Second Goal: Load Shifting}

The marginal cost of electricity at peak load, taking both production
and distribution into account is considerable. It is desirable to
reduce peak loads by shifting usage from high load to low load times
over a 24-hour period or even within a shorter period. Reducing peak
loads lessens the risk of overburdening the system, reduces the dependence on marginal electrity production systems -- often with more noticeable environmental impact and with greater production cost per unit, and allows the
power grid to be specified for a more rational capacity utilisation.

\subsection{Third Goal: Preserving the Standard of Living}\label{std}

An immediate political concern is to accomplish a lowered electrical energy
turnover and a more balanced load over time, and to do this without
impacting negatively on living standards. The savings should be
predominantly directed at unconscious spill and waste rather than at
reducing quality of life. 

In addition to this, the most immediate and pressing need of many
users is expressed in terms of life management and coping -- we do not
wish to add burdens to the harried everyday life of consumers.
Our goal is to help the individual household consumer keep electrical energy
usage an appropriately small part of everyday life, afford
the user higher awareness, better sense of control, without
intruding on the general make-up of the home by introducing new,
cumbersome, and unaesthetic devices.

\section{Studies of Users}

Feedback for better awareness or control of energy usage is well
studied. A literature review from the University of
Oxford~\cite{oxford} indicates savings between 5-15\% from direct
feedback, and also states that ``... time-of-use or real-time pricing
may become important as part of more sophisticated load management and
as more distributed generation comes on stream.''. Much of the work
listed in the review concerns feedback in the form of informative
billing and various types of displays (even ambient ones), with the
aim of providing users with a better understanding of their energy
usage. {\em Smart metering} is a general approach to building better and more intelligent meters to monitor electricity usage, to raise awareness among consumers, and to provide mechanisms of control either to distributors or consumers: modern electricity meters are frequently built to be sensitive to load balancing issues or tariff variation and can be controlled either through customer configuration or through distributor overrides beyond the control of the consumer \cite{owenward,lund1,lund2}. Better metering has a potential impact on energy load peak reduction and the allows for the possibility for time-of-use pricing, issues which have influenced the design of the tea kettle prototype presented later in this paper. 

Furthermore, initiatives such as UK's national Design Councils top
policy recommendations from work with users, policy makers, and
energy experts, highlight user awareness and control given by e.g. more
detailed real-time monitoring of appliance energy usage, controlled
through an ``allowance'', household ``energy collaboratives'' and energy
trading clubs \cite{ukdc}.

Our project shares goals with smart metering projects but focusses
more on the control aspect than most other approaches --- on how to
allow users to influence their use of energy in the home environment.

From exploratory interview studies performed in homes and households
by ourselves during the prototype design phase of our project, we have
found that consumers in general are quite interested in taking control
of their energy turnover: they feel they understand little of it and
that the configuration and control of their home appliances are less
in their hand than the consumers would like them to be. Our subjects
in several of the at-home interviews we performed expressed puzzlement
and lack of control in face of incomprehensible and information-dense
energy bills -- none wished to receive more information on the bill
itself. For instance, interview subjects were not comfortably aware of
the relation between billing units ($kWh$) and electricity usage in
the household.

However, several of our subjects had instituted routines or behaviour
to better understand or monitor their energy usage: making periodic
notes of the electricity meter figures, admonishing family members to
turn off various appliances, limiting the usage of appliances felt to
be wasteful of energy, switching energy suppliers to track the least
costly rates offered on the market. Many of these routines were
ill-informed or ineffectual (e.g. turning off lighting, while not
addressing other appliances such as home entertainment systems), and
were likely to have little or no effect on the actual electricity
usage or cost, but it afforded the users in question some sense of
control.

Wishing to gain better control is partially a question of home
economics: rising electricity bills motivates consumers to be more
aware of electricity usage and to economise. However, while economy
alone cannot become a strong enough motivating factor (as noted above, for reasons
extraneous to the energy and environment field),
consumer behaviour can be guided by other indicators. Societal
pressure to behave responsibly accounts for much of consumer
behaviour, e.g. in the form of willingness to participate in a more
sustainable lifestyle --- witness the high rate of return of drink
containers in spite of the low remuneration given by the deposit
system. \footnote{Close to ninety per cent of drink containers are
  returned by consumers in Sweden, with many European jurisdictions
  showing similar figures, according to the Swedish Waste Management
  trade association (www.avfallsverige.se). The presence of a deposit
  system seems to have a strong effect on the willingness to make the
  effort to return containers, even while the deposit usually is low
  and does not translate into adequate remuneration for the time and
  effort spent.} A companion motivator can be found in a general
pursuit better to be able to retain the initiative and control over an
increasingly complex set of systems in the household.

\section{Guiding Principles}

To empower users to take control of their household energy turnover,
we work to harness the interest consumers show towards the electrical energy usage
issue. Our conviction is that to harness this interest we need to
design and develop systems that are effortless, fun, and effective to
use. We wish to develop pleasurable aspects of new technology --- and we do
not wish to build systems to use value statements to influence or
distress its users. To achieve these goals we suggest a set of guiding principles to
follow. These principles are effectivisation, avoiding drastic
lifestyle changes, utilizing ambient and physical interfaces,
providing comparison mechanism, make systems socially aware and
provide both immediate and overview feedback.  In the following
section we will motivate our choice of principles.

\subsection*{Work towards effectivisation, avoiding drastic lifestyle changes}

As discussed in section~\ref{std}, we do not wish to burden the consumer with further tasks or cognitive demands, nor to lower the standard of living in households. Rather then reducing quality of life we need to work towards more effective use of energy in general, and of electric energy in this specific case: effective in both the sense of higher efficiency but also correctly placed in time and space. In doing so we shall not enforce users to introduce drastic life style changes which would become another burden for them. 

\subsection*{Use ambient interfaces and physical interfaces}

As mentioned above an important aspect of designing artifacts for the home and the household is to avoid drastic life style changes, i.e. not disrupting the behavioural patterns of the inhabitants. Furthermore it is not desired to ruin the aesthetic qualities of the interior. Ambient interfaces \cite{kalleny} share the design principle to not disrupt behavioural patterns and thus the use of ambient interfaces suits very well to be employed. Furthermore utilization of physical interfaces allows us to design interaction that embed into the aesthetics of the artifact and that the actual interaction with the artifact is not drastically changed. \cite{kallekanske}

\subsection*{Use comparison mechanisms}
To reduce energy usage we must slowly change the user's behaviours. As a basis for any system for behavioural modification, we must provide a benchmark or measure with which to compare individual (in our case, typically aggregated by household) performance. We will want to provide comparison mechanisms to give the consumer a sense of change, both as regards overall lifestyle and for individual actions which influence energy turnover in the home. In our current set of prototypes, we explore comparison over time ("Am I doing better than last week") and over peer groups (using a recommendation system framework \cite{KarlgrenJ:94,sicsprint417}). 

\subsection*{Build socially aware systems}
Energy usage and related behaviour, as intimated above, is not only an individual question. Use aggregated over a large number of households will have greater impact both on general usage levels and on load balance than will individual usage profiles; in addition, the effects of social pressure, both to conform and to lead, can be harnessed to make a difference. Therefore it is important to create designs that operationalise some of the social context that normally is less visible in household situations. If this is done in a manner which avoids pitting users against each other in individual contests of appropriate behaviour, and which does not succumb to pointing fingers at those who do not conform to norms, such a social aware system will provide users to not only compare or learn from their own usage but others as well.

\subsection*{Immediate feedback and overview feedback}
While the learning which is required by the consumer is on a fairly high cognitive level, and is intended to increase awareness and contribute to a sense of empowerment and control, it should be based on standard learning mechanisms. It is well established that in any learning situation, the timing of the stimulus must be in relevant juxtaposition with the contigent action --- the highlevel behaviour or individual actions the consumer is engaged in. \cite{KirschEA:04} The stimuli or signals to provide
as feedback to the user must be {\em appropriate} and have {\em
  informational value}  --- in our case, they must not be overloaded with other information and not clash with other signals or messages the user may be interested in, and it must also have relevant cognitive content so as not to deteriorate into mere background noise.
Specifically, we will keep separate the immediate feedback necessary to learn from actions from the overview sense necessary to be aware of lifestyle effects.

\subsection*{Plan and build for added benefits}
The economical benefits provided through higher energy awareness is not enough to catch the eye of consumers, but must include other benefits and thus be designed as part of a larger platform. One possible added benefit is to provide added control and to empower the indvidual consumer and emergent groups of consumers. A possible integration of energy issues with e.g. safety monitors, time management services, and communication systems might be one solution.

\section{Prototype Example -- The Socially Aware Tea Kettle}

Electric stand-alone tea kettles are the preferred device for heating
water if the alternative is using a pot on an electric stove: tea
kettles are much more energy efficient. To contribute to the
overriding goal of reducing electricity usage it is thus useful to
encourage households to move from stove-top saucepans to stand-alone
kettles.

However, kettles occasion usage peaks: they use power up to 2 kW
and are among those home appliances which require the highest current
when switched on. This is observable by the consumer, in that
switching on a tea kettle for many consumers causes a minor but visible brownout in
the home, dimming lights momentarily. In addition, kettle usage
patterns are cyclic. People have more or less the same diurnal
behaviour patterns and tend to heat water for hot beverages and for
other cooking purposes at about the same times.

The tea kettle is thus a useful and illustrative example of household
electricity appliance: it has a high wattage and its usage is
non-random with similar usage patterns across households.

As noted above, reducing peak loads by shifting usage is a desirable
goal. To this end, to illustrate the possibilities inherent in
aggregating households into cooperating pools of users for
load-balancing purposes, we have built a socially aware tea kettle,
suitable for pooling usage across several households. Pooling usage and reducing peak loads allows an aggregation of households the potential to negotiate lower average rates (possibly offset by higher peak load rates at the margin). In keeping with the guiding principles given above, we do not wish to burden the consumer with calculations or other cognitively demanding operations at the point in time where their focus is on producing hot beverages. Instead of providing numerical or graphical information, deflecting the attention of the consumer from tea to tariffs and time schedules, we provide an enhanced tool whose primary purpose remains heating water.

\begin{figure}[htbp]
\begin{center}
\includegraphics[ height=7cm]{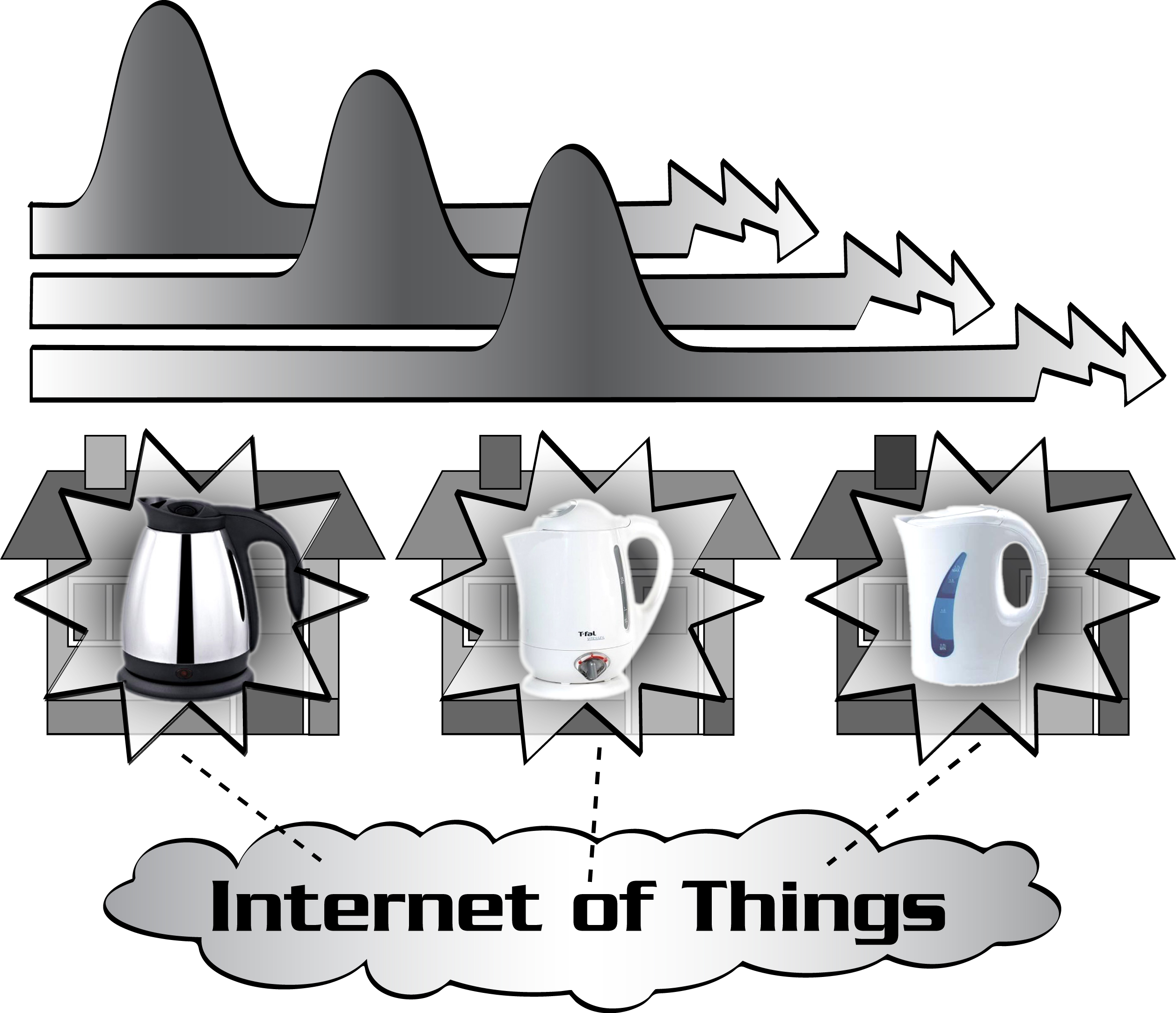} 
\end{center}
\caption{Pooling usage to achieve load balance.}\label{pool}
\end{figure}

The underlying assumption is that if a number of households with
suitably dissimilar habits are pooled, their power requirement can be
balanced over a time period, if less time-critical energy requirements
at times can be shifted from the immediate moment they are ordered to
some less loaded point in time. A typical example of pooling currently
under experimental deployment in households is delaying hot-water cisterns from immediately
replacing the hot water used by morning baths and showers. If
hot-water cisterns are programmed to heat the water when the overall
network load is low, the aggregate cost of heating water can be kept down~\cite{lund1,lund2}.
Figure~\ref{pool} illustrates how the power usage of various electrical
appliances (examplied by tea kettles) can be spread out in time to
acheive a better load balance if the households are pooled. 

Similarly, we posit that heating water is only partially
time-critical: when someone orders up a pot of boiling water, quite
frequently the hot water can be delivered within a flexible time
window rather than immediately. If a kettle can be made aware of the
current and near-future load among a number of pooled households, it
can inform its user of when within the next few minutes a kettleful
water can be heated at lowest strain to the delivery grid (and,
presumably, at lowest cost, if price is used as an additional signal
to consumers).

\begin{figure}[htbp]
\begin{center}
\includegraphics[height=8cm]{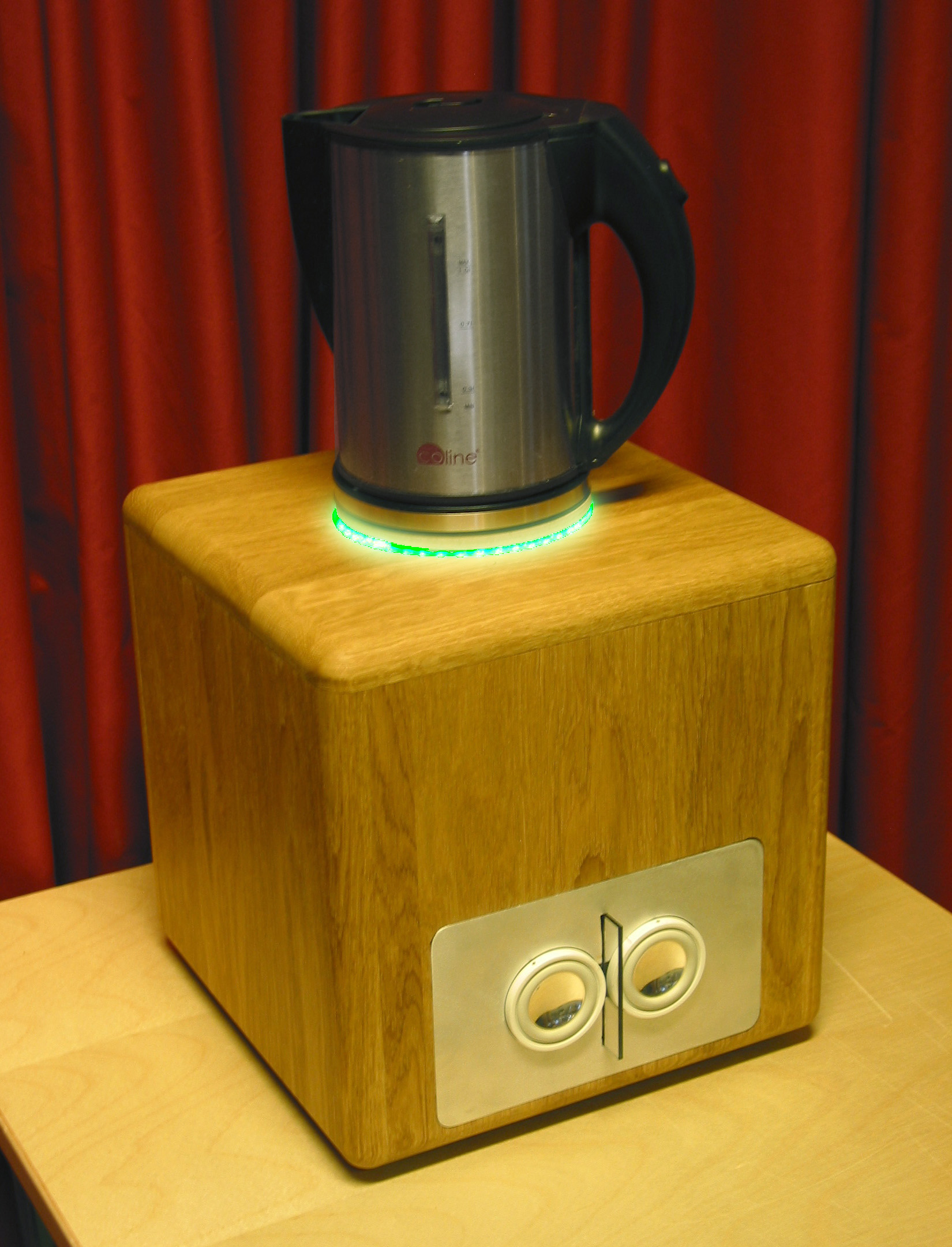} 
\caption{Current kettle prototype. A production unit will have the same form factor as a standard off the shelf tea kettle.}\label{prototype}
\end{center}
\end{figure}

Our example fully-functional kettle is mounted on a rotating base, and
provides the user with rotational force feed-back to indicate
booked capacity in a window over the next few minutes. A picture of the current prototype is shown in Figure~\ref{prototype}. This allows the
user to rotate the kettle to get a sense of when the load is light
(less force is needed) or heavy (the kettle resists being rotated into
that position). The user is always able to override the recommended
time slot, and select immediate delivery or non-peak time.

\subsection{Technology}

The wooden cabinet on top of which the tea kettle is placed contains
various hardware devices that are necessary for its operation. At the
heart of this setup is a Java microcontroller responsible for Internet
communication as well as measurement and control functionality.  Using
a clamp meter connected through i2c interface, the microcontroller
monitors electric current used by the kettle. An i2c connected 240V
power switch is incorporated to allow the processor to start and stop
the tea kettle at designated times. The prototype also uses a modified
force feedback steering wheel, modified from an off-the-shelf car game
controller, connected to the base of the kettle to yield the
friction-based rotation of the kettle. In addition, there is an array
of LEDs positioned around the base of the kettle used to visually
convey information such as active bookings. 

\begin{figure*}[htbp]
\begin{center}
\includegraphics[ height=7cm]{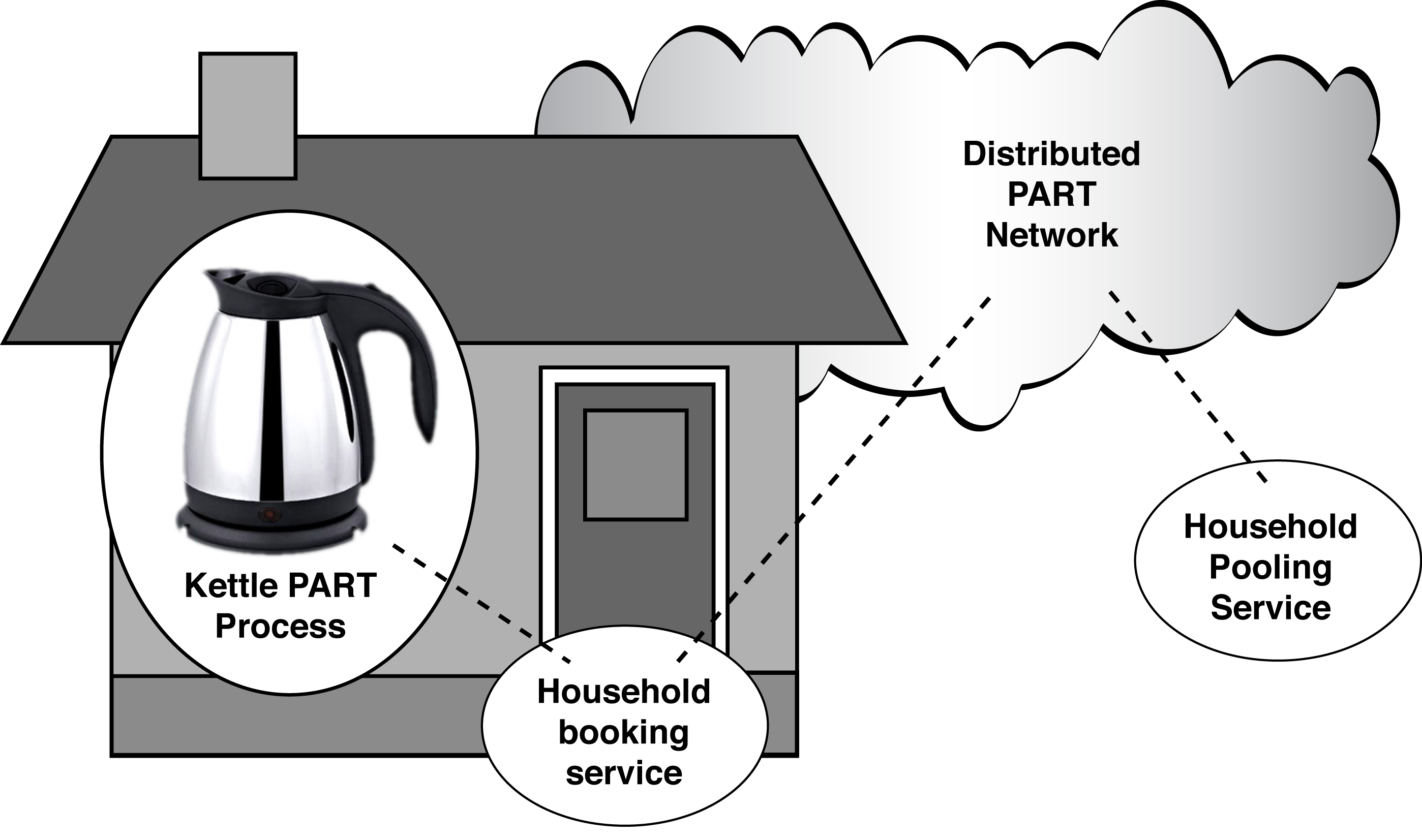} 
\end{center}
\caption{System overview.}\label{part}
\end{figure*}

The microcontroller runs a kettle software process which controls the
hardware devices inside the wooden cabinet (forced feedback device,
power switch, etc.), and also communicates with a household booking
service over IP. The booking service is connected to all household
electrical appliances, and monitors their current as well as predicted
future power usage. The booking service is in turn connected and
reports to a pooling service, which supervises the activity over many
households. The pooling service is thus able to construct an
aggregated ``profile'' of the predicted near-future power consumption of
all connected households. This profile is then continuously
communicated back to the booking services, and finally the individual
appliances (e.g., the tea kettle prototype), where it is used to
convey operating conditions that the user may wish to consider. The
profile may be biased by any changes in the power suppliers future
cost levels or by maximum power usage limits that may be stipulated in
contracts between the power supplier and the pooled households. In the
tea kettle prototype the profile is used to control the force
feed-back mechanism acting on the kettle. The friction at a certain
angle of rotation represents the predicted load a number of minutes
into the future. The system architecture is illustrated in Figure~\ref{part}. The kettle software control process, as well as the booking and
pooling services, are developed using PART~\cite{part}. PART is a light-weight
middleware, written in Java, for developing pervasive applications
that run on a range of different devices. PART is available as open
source, under BSD license, and can be downloaded at
http://part.sf.net.

The kettle can be rotated approximately 360 degrees, which currently is set to correspond
to fifteen minutes of time. At one end of the interval
(starting from degree 0), the friction feedback to the user indicates bookings
and thus the predicted load in the immediate
near future, while at the other end (around degree 360), the user can sense the predicted load fifteen minutes further along in time. When the user has found a suitable position, the kettle is
activated by pressing the ordinary ``on'' button. This event is caught
by the kettle controller hardware, which will temporarily switch off the
kettle's power. The power will then be switched back on by the controller
at the time corresponding to the kettle rotation, which will cause the water to
start heating.

The reason why the rotation interval is fairly short (minutes rather
than hours) is that we believe that roughly fifteen minutes is as long
as tea and coffee drinkers can be expected to wait for their hot water. The assumption
is that users want their hot water now, but that they might be
prepared to wait a while in order to getter a better price, or to help
spread the load among their peers. We have considered it unlikely that
the user would want to schedule the heating of water for cooking or
hot beverages hours into the future. One implication of this is that
we also believe that once the user starts to turn the kettle, the
chances are high that a booking will actually be made (the user will want
the hot water now!), even if the load turns out to be high in the entire 
fifteen minute interval. This has allowed us to keep the
interface fairly simple, based more or less entirely on the force
feedback mechanism, and avoiding displays, buttons, etc, which might
would have been required to support a more complex booking procedure
(e.g., allowing the user to set constraints on price, time, load,
etc.) All of these assumptions have to be confirmed via user studies,
which are planned for the near future.

\subsection{The Home Configuration Panel}

Once the previous prototype is in place, the integration of various
home services will hinge on getting an overview of the home
configuration. How this can be achieved without overwhelming the user
is a daunting design task, and this interface is only in planning
stages as of yet -- but it will involve a careful combination of well
designed controls with defaults, examples, and preferential
selections in face of difficult choice. \cite{atta,nio}

We foresee in this case an integration of energy issues with
e.g. safety monitors, time management services, and communication
systems.

\subsection{Community Tool}

Similarly to the home configuration panel, an overview and control
system is needed for the pooled community household power
requirements. This could involve social awareness mechanisms as well
as an opportunity to test different business models towards the
provider. Current implementation consists of a graph visualizing web
browser applet communicating to the rest of the system via a servlet.

\section{Conclusions}

This prototype is designed to illustrate three
strands of thought. It works towards an increased {\em social
  awareness} in the household; it is based on an informed design of
{\em immediate feedback} of aggregate information; it is an example
of {\em unobtrusive design} to fit in the
household without marring its home-like qualities, and as a platform and base for 
further design work, it combines our
design principles in a fully functional prototype.

The kettle is an instantiation of our design principles, but by no
means the only possible one, and is not intended to be deployed in the
consumer market as is -- it serves as an illustration of how social
factors are or can be a parameter in our behaviour even within our
homes. The underlying principle of an internet of household appliances
can be instantiated in many ways, of which this is but one: energy
usage is not a prototypically social activity, but in aggregating the
behaviour of many peers into a pool of usage it becomes more social
than before. The point of this prototype is that with a tangible yet
unobtrusive interface we are able to provide users with a sense of
social context and superimpose a behavioural overlay over individual
action to aggregate usage patterns into a whole which is more
functional from a societal perspective. We will be exploring this
avenue of social technology in further prototypes, with other design
realisations of the same principles.

Our tea kettle prototype illustrates the importance of load balancing and of
user control. However, it also raises the requirement of providing
users with immediate feedback directly connected to action -- a
well-established principle in the study of learning. As further
developments of that principle we are field testing variants of electricity
meters which provide both overview information of aggregated weekly usage and
in other instances reacts to user action by giving auditory and visible
feedback when an appliance is connected to the net.

Our tea kettle prototype illustrates the possibility of designing an interface which both
provides immediate lean-forward information for the action the user is about to
engage in, but without requiring cognitive effort -- the information is encoded
in rotation and visual feedback which needs little or no parsing. The attention of the user is not deflected from the primary goal at hand; the tool remains a tool designed for its primary function.

\section{Acknowledgements}

The research reported in this paper is part of and sponsored by the
Swedish Energy Agency research program ``IT, design and energy'', project
ERG, project number 30150-1. The ERG project adresses consumer
awareness, understanding, and control of the energy usage of tomorrow's
household.

\bibliographystyle{plainnat}
\bibliography{tekokare-shared}  

\end{document}